# A Laboratory log(*gf*) Measurement of the Ti II 15873.84 Å *H*-band Line in Support of SDSS-III APOGEE

M. P. Wood[1], J. E. Lawler[1], and M. D. Shetrone[2]

[1]Department of Physics, University of Wisconsin-Madison, Madison, WI 53703 USA;

mpwood@wisc.edu, jelawler@wisc.edu

[2]McDonald Observatory, University of Texas at Austin, Fort Davis, TX 79734 USA;

shetrone@astro.as.utexas.edu

Abstract

The SDSS-III APOGEE collaboration has identified a single useable line in the *H*-band spectra of APOGEE target stars arising from a singly ionized species. This line of Ti II ($\lambda_{air}$ = 15873.84 Å) is therefore of great importance for use in stellar surface gravity, or log($g$), determinations via the Saha equation. While a theoretical estimate of the line strength exists, to date no laboratory measurement of the line strength has been reported. Herein we report an absolute laboratory transition probability measurement for this important Ti II line. A relative line strength measurement is made of the Ti II *H*-band line of interest and a reference line with a previously reported absolute transition probability. This ratio is measured using multiple spectra of a high-current water-cooled HC lamp recorded with a calibrated FT-IR spectrometer.

1. Introduction

The increasing availability and performance of mercury cadmium telluride (HgCdTe) Astronomical Wide Area Infrared (IR) Imager (HAWAII) arrays is revolutionizing IR astronomy (e.g., Hodapp et al. 1994, Ives & Bezawada 2007). Such arrays are critical to the much anticipated NASA *James Webb Space Telescope*, which is expected to provide extraordinary new high redshift observations and revolutionize our understanding of the early Universe. These arrays are also critical to the ground based Sloan Digital Sky Survey-III (SDSS-III) Apache Point Observatory Galactic Evolution Experiment (APOGEE; see section 4 of Eisenstein et al. 2011). These arrays, combined with a powerful new cryogenic spectrometer (Wilson et al. 2010, Wilson et al. 2012), are providing chemical abundances for a large number (~ 100,000) of stars in the Galactic bulge, bar, disk, and halo. By operating at IR wavelengths in the *H*-band (~ 1.5 - 1.7 μm), APOGEE can study stars in parts of the Galaxy, especially the bulge, that are not accessible in the optical region due to interstellar dust extinction. The above two examples represent current high profile projects, but they are just the leading edge of a great era for IR astronomy, e.g. the Multi-Object Optical and Near-Infrared Spectrograph on the VLT (Cirasuolo et al. 2011).

Although the transparency of interstellar dust at *H*-band wavelengths is a huge advantage for APOGEE, working in the rather small wavelength interval of the *H*-band ($\Delta\lambda \sim 1800$ Å) poses some challenges. Many high resolution abundance analyses of field stars determine stellar surface gravity, or $\log(g)$, using an ionization balance from the Saha equation (perhaps with a correction). This requires that detectable neutral, or first spectra, and singly ionized, or second spectra, features are present in the spectra. Typical red giant stars of the APOGEE study have lines of about 15 elements in the *H*-band, including the most common metals (C, N, O), some α-elements, various Fe-group elements, and two odd-Z elements, Na and Al. Most of the observed spectral features are from first spectra lines, such as Fe I, in addition to significant contributions from diatomic molecules. The shortage of second spectra lines in the IR is not surprising and the APOGEE team developed some techniques to determine stellar parameters without using ionization balance (Garcia Perez et al., in prep). However, we recently identified a fairly clean Ti II feature which, when combined with previously known lines of Ti I, could be used in $\log(g)$ determinations. This line was previously overlooked because the wavelength in the Kurucz database (Kurucz & Bell 1995) is 0.2 Å off from the observed feature at $\lambda_{air}$ = 15873.84 Å. In the recent study of Ti II transition probabilities by Wood et al. (2013), this Ti II line had been observed. That study benefitted from recently reported Ti II energy level measurements (Saloman 2012), and using these new energy level data removed the 0.2 Å discrepancy on the Ti II *H*-band line. However, the line was dropped from the final published list of 364 log(*gf*) values as available data on the Ti II line of interest and on many other weak lines did not have satisfactory signal-to-noise (S/N) ratios for inclusion in the final published results. Such lines are kept in the normalization of branching fractions, but because they are so weak they have very little effect on the transition probabilities and negligible effect on uncertainties of transition probabilities for stronger lines.

The theoretical log(*gf*) of the Ti II line (log(*gf*) = -1.925) in the Kurucz database (Kurucz & Bell 1995) yields a feature of approximately the correct strength to fit the high resolution solar and Arcturus atlases (Hinkle et al. 1995), although offset by 0.2 Å. While agreement in line strength between the observed spectral feature and theoretical calculations is encouraging, a laboratory measurement is greatly needed to improve confidence in the line strength value. In addition, the Kurucz value does not include an uncertainty estimate, which makes exploring possible NLTE effects and understanding sources of error in derived log($g$) difficult. Herein we report an absolute laboratory measurement of log($gf$) = -1.90 ± 0.08 for the Ti II $H$-band line at $\lambda_{air}$ = 15873.84 Å.

2. Experiment

A laboratory measurement is made of the relative intensity of the Ti II $H$-band line of interest ($\lambda_{air}$ = 15873.84 Å) and a reference line ($\lambda_{air}$ = 9432.151 Å) with a previously reported absolute atomic transition probability (Wood et al. 2013). Both transitions connect to the $3d^2(^3F)4p\ z^2F°_{7/2}$ upper level at 31490.912 cm$^{-1}$, with the $H$-band line of interest connecting to the $3d4s^2\ c^2D_{5/2}$ lower level at 25192.965 cm$^{-1}$ and the reference line connecting to the $3d^3\ b^2F_{7/2}$ lower level at 20891.790 cm$^{-1}$. The reference line was put on an absolute scale using a laser-induced fluorescence (LIF) lifetime measurement (Bizzarri et al. 1993). From the relative intensity measurement reported in this study, an absolute transition probability for the Ti II $H$-band line is determined.

2.1. Experimental Setup

Laboratory measurements of the line strength for both the Ti II $H$-band line of interest and the previously reported reference line (Wood et al. 2013) are made using a Thermo Nicolet Nexus 870 Fourier transform-infrared (FT-IR) spectrometer. Emission spectra are obtained from a custom water-cooled hollow-cathode (HC) discharge lamp. Relevant parameters for both the setup of the FT-IR and the HC lamp for each measurement are given in Table 1. The HC lamp consists of a copper cathode with a Grade 4 titanium insert (inner diameter ~ 1 mm) providing a sputtering surface. Both of these Ti II lines are very weak, with branching fractions << 0.01, so the use of a water-cooled HC lamp running at high current is essential to achieve the necessary density of singly ionized Ti atoms to record emission spectra of these lines. The weakness of these branches eliminates any concerns regarding optical depth errors. Even with the high-current spectra, satisfactory S/N is not achievable without a reduction of the inherent multiplex noise of the FT-IR through the use of filters. Multiplex noise, in which the quantum statistical (Poisson) noise from all spectra features, particularly strong visible and IR features, is evenly redistributed throughout the entire spectrum, is a significant disadvantage when attempting to measure these weak IR lines. To reduce the multiplex noise, filters are used to block all wavelengths except for a narrow region centered on each line of interest. The filters employed are a narrow-band Melles-Griot 950 interference filter tipped at ~ 12° (03FII016, $\lambda_{peak} \approx 9430$ Å, FWHM ≈ 130 Å) for the previously measured reference line and the broad-band ThrillCam *H*-band filter[1] ($\lambda_{mid} \approx 16400$ Å, FWHM ≈ 3150 Å) for the *H*-band line of interest.

Figure 1 shows the FT-IR spectrum of the reference line ($\lambda_{air}$ = 9432.151 Å) previously reported by Wood et al. (2013), while Figure 2 shows the FT-IR spectrum of the *H*-band line ($\lambda_{air}$ = 15873.84 Å) of interest in this study. Both lines are from the same data series (Index 10 in Table 1) and are chosen to be representative of the S/N achieved using the high-current HC lamp and FT-IR with filters to suppress multiplex noise. The intensity units are arbitrary since the final measurement of interest is the relative intensity ratio of the two lines, which is then used to determine an absolute atomic transition probability for the Ti II *H*-band line. Integrated line intensities are determined by fitting Gaussian profiles to the data using a nonlinear least squares (NLLS) technique. These weak lines, which lack isotopic or hyperfine structure in our spectra, are well fit using this technique, which also allows for a clean separation of the Ti I blend shown in Figure 1.

---

[1] On loan from the Department of Astronomy at the University of Virginia.

## 2.2. Radiometric Calibration

A relative radiometric calibration of the FT-IR data described above is needed to tie the atomic transition probability of the *H*-band line to the reference transition probability normalized using accurate, absolute radiative lifetimes measured via LIF (Bizzarri et al. 1993). The selected $\lambda_{air}$ = 9432.151 Å reference line for this work is, of course, from the same upper level at 31490.918 cm$^{-1}$ and has an Einstein $A = (8.4 \pm 1.4) \times 10^4$ s$^{-1}$ and log($gf$) = -2.05 (Wood et al. 2013). The majority of the uncertainty on this measurement is due to systematic uncertainty in the calibration over such a large wavelength interval, from the dominant near-UV branch of the upper level to the near-IR reference line used in this work.

The separation of these two lines, which is almost a factor of two in wavelength, requires a careful relative radiometric calibration. The same narrowband filters used to suppress multiplex noise in the FT-IR data are used during the calibration with a NIST-traceable E$^2$T Blackbody (cavity) Calibrator operating at a temperature T = 1673 K. The spectral radiance in units of photons per unit time per unit wavenumber per unit area per solid angle from the blackbody source is

$$I_{BB}(T) = \frac{2c\sigma^2}{\exp(hc\sigma/k_B T) - 1}$$

where $\sigma = 1/\lambda$ is the wavenumber, $c$ is the speed of light, $h$ is Planck's constant, and $k_B$ is the Boltzmann constant. We remind the reader that wavenumber units are the natural and preferred units of interferometric spectrometry and are always given in vacuum, whereas our FT-IR is operating in air. Air wavelengths for the two lines measured in this study are calculated from the updated Ti II energy levels (Saloman 2012) using the standard index of air (Peck & Reeder

1972). The ratio of the Einstein $A$ coefficients for the $H$-band line (15874 Å) to that of the reference line (9432 Å) is

$$\frac{A_{15874}}{A_{9432}} = \frac{L_{15874}}{L_{9432}} \frac{B_{9432}}{B_{15874}} \left(\frac{9432.1508}{15873.839}\right)^2 \frac{\exp(hc\sigma_{9432}/k_B T)-1}{\exp(hc\sigma_{15874}/k_B T)-1}$$

where $L$ is the wavenumber integrated line signal strength from the FT-IR spectra and $B$ is the wavenumber integrated blackbody continuum signal strength from the FT-IR calibration spectra. A rotatable mirror couples the $E^2T$ Blackbody Calibrator into the FT-IR to allow calibration spectra to be measured for each of the emission spectra listed in Table 1. The only setting changed between the emission and calibration spectra is the resolution. A broader limit of resolution (0.964 cm$^{-1}$) is chosen for the calibration spectra, both because of the smooth structure of the blackbody continuum and in order to reduce the collection time. The other FT-IR settings (including wavenumber intervals and filters) remain the same in order to disturb the FT-IR as little as possible and obtain an accurate relative radiometric calibration. Unlike the HC emission spectra, the blackbody calibration spectra have been background corrected. The $B$ integrations each cover the same interval (~ 10 cm$^{-1}$) centered on the wavenumber of the two lines. Owing to the lack of structure, and unlike the line strength integrations, the $B$ integrations are found using a simple numerical integration routine. Based on the repeatability in the wavenumber integrated continuum signal strength, the $E^2T$ Blackbody Calibrator shows ~ 2% stability over several hours of operation.

## 3. Result

The final relative line strength ratio ($A_{15874}/A_{9432}$) is 0.501 ± 0.045, based on 13 separate measurements from the spectra listed in Table 1. This represents the weighted mean and weighted standard deviation of the line ratios calculated for each of the 13 indexed measurements in Table 1. The random error attributable to the weighted standard deviation is quoted, as this is larger than the weighted uncertainty on the measurement. Sources of uncertainty contributing to the weighting factors are shown in Table 2. Combining this result with the previously reported absolute transition probability for the 9432.151 Å reference line (Wood et al. 2013) provides an absolute transition probability for the *H*-band line of interest: $A_{15874} = (4.2 \pm 0.8) \times 10^4$ s$^{-1}$. The uncertainty on this measurement is the line strength ratio uncertainty and the reported 9432.151 Å transition probability uncertainty combined in quadrature. As shown in Table 2, the 9432.151 Å transition probability uncertainty is the dominant source of uncertainty for this measurement. The following equation converts the transition probability to a log(*gf*) value:

$$\log_{10}(gf) = \log_{10}\left(g_u A \lambda^2 \cdot 1.499 \times 10^{-14}\right)$$

where $g_u$ represent the degeneracy of the upper level (2*J* + 1; $g_u$ = 8 for this level), *A* represents the transition probability, and $\lambda$ is the transition wavelength in nm (Thorne et al. 1999). From this, the log(*gf*) value for the Ti II *H*-band line of interest at $\lambda_{air}$ = 15873.84 Å is determined to be log(*gf*) = -1.90 ± 0.08.

## 4. Conclusion

A laboratory measurement of the log(*gf*) value for the *H*-band Ti II line at $\lambda_{air}$ = 15873.84 Å is reported. This line represents the only singly ionized transition in the *H*-band observed by the APOGEE collaboration, and is therefore of great importance for accurate gravity measurements in the photospheres of APOGEE target stars. Using a FT-IR spectrometer, a measurement of the relative line strength ratio between this *H*-band line and a reference line with a previously reported absolute transition probability measurement is performed. A NIST-traceable blackbody source provides an accurate relative radiometric calibration of the FT-IR. Filters and a high-current water-cooled HC lamp are used to achieve satisfactory S/N for the line intensity measurements.

The authors would like to acknowledge the contribution of J. Alter during early data collection.  The authors also thank S. Majewski of the University of Virginia for loaning the *H*-band filter used during this study.  This work is supported by NSF grant AST-1211055.

Figure Captions

Figure 1. FT-IR HC emission spectra of the previously measured Ti II reference line at $\sigma$ = 10599.128 cm$^{-1}$ ($\lambda_{air}$ = 9432.151 Å). The line to the right is a Ti I line at $\sigma$ = 10599.728 cm$^{-1}$. The highest resolution available with the FT-IR allows these lines to be almost completely resolved.

Figure 2. FT-IR HC emission spectra of the Ti II *H*-band line of interest at $\sigma$ = 6297.953 cm$^{-1}$ ($\lambda_{air}$ = 15873.84 Å). The sloping background is due to thermal emission from the Ti cathode insert.

Table 1. FT-IR spectrometer and HC lamp parameters for each of the 13 spectra used in this study. A SiO$_2$ beamsplitter and thermo-electrically cooled (TEC) InGaAs detector provide the best near-IR performance for the FT-IR, and are thus used for all spectra listed. The water-cooled HC lamps is operated with Ar buffer gas.

| Index | Date | Buffer Gas Pressure (Torr) | Lamp Current (mA) | Serial Number[a] | Wavenumber Range (cm$^{-1}$) | Limit of Resolution (cm$^{-1}$) | Coadds | Filter[b] |
|---|---|---|---|---|---|---|---|---|
| 1 | 2013 Dec. 18 | 9 | 450 | 13L18#03 | 10550 – 10650 | 0.060 | 100 | MG 950 |
|   |   |   |   | 13L18#04 | 6250 – 6350 | 0.060 | 100 | *H*-band |
| 2 | 2014 Jan. 2 | 10 | 550 | 14A02#03 | 10550 – 10650 | 0.060 | 80 | MG 950 |
|   |   |   |   | 14A02#04 | 6250 – 6350 | 0.121 | 160 | *H*-band |
| 3 | 2014 Jan. 2 | 15 | 650 | 14A02#09 | 10550 – 10650 | 0.060 | 80 | MG 950 |
|   |   |   |   | 14A02#10 | 6250 – 6350 | 0.121 | 160 | *H*-band |
| 4 | 2014 Jan. 2 | 20 | 750 | 14A02#15 | 10550 – 10650 | 0.060 | 100 | MG 950 |
|   |   |   |   | 14A02#16 | 6250 – 6350 | 0.060 | 100 | *H*-band |
| 5 | 2014 Feb. 12 | 12 | 550 | 14B12#03 | 10550 – 10650 | 0.060 | 100 | MG 950 |
|   |   |   |   | 14B12#04 | 6250 – 6350 | 0.060 | 100 | *H*-band |
| 6 | 2014 Feb. 12 | 12 | 450 | 14B12#09 | 10550 – 10650 | 0.060 | 100 | MG 950 |
|   |   |   |   | 14B12#10 | 6250 – 6350 | 0.060 | 100 | *H*-band |
| 7 | 2014 Feb. 12 | 12 | 550 | 14B12#15 | 10550 – 10650 | 0.060 | 100 | MG 950 |
|   |   |   |   | 14B12#16 | 6250 – 6350 | 0.060 | 100 | *H*-band |
| 8 | 2014 Feb. 13 | 15 | 450 | 14B13#03 | 10550 – 10650 | 0.060 | 100 | MG 950 |
|   |   |   |   | 14B13#04 | 6250 – 6350 | 0.060 | 100 | *H*-band |
| 9 | 2014 Feb. 13 | 15 | 550 | 14B13#09 | 10550 – 10650 | 0.060 | 100 | MG 950 |
|   |   |   |   | 14B13#10 | 6250 – 6350 | 0.060 | 100 | *H*-band |

| | | | | 14B13#15 | 10550 – 10650 | 0.060 | 100 | MG 950 |
|---|---|---|---|---|---|---|---|---|
| 10 | 2014 Feb. 13 | 15 | 550 | 14B13#16 | 6250 – 6350 | 0.060 | 100 | *H*-band |
| 11 | 2014 Feb. 13 | 15 | 650 | 14B13#21 | 10550 – 10650 | 0.060 | 100 | MG 950 |
| | | | | 14B13#22 | 6250 – 6350 | 0.060 | 100 | *H*-band |
| 12 | 2014 Feb. 17 | 12 | 550 | 14B17#03 | 10550 – 10650 | 0.060 | 100 | MG 950 |
| | | | | 14B17#04 | 6250 – 6350 | 0.060 | 100 | *H*-band |
| 13 | 2014 Feb. 17 | 12 | 650 | 14B17#09 | 10550 – 10650 | 0.060 | 100 | MG 950 |
| | | | | 14B17#10 | 6250 – 6350 | 0.060 | 100 | *H*-band |

[a]Each indexed measurement comprises two individual spectra, one each for the reference line and the *H*-band line of interest.

[b]Filters include a narrow-band Melles-Griot 03FII016 (MG 950) interference filter, tipped at ~ 12° to shift the band-pass to shorter wavelengths, and the broad-band ThrillCam *H*-band filter.

Table 2. Sources of uncertainty contributing to the final transition probability uncertainty for the Ti II *H*-band line at 15873.84 Å.

| Uncertainty | Source | Contribution |
|---|---|---|
| Line Strength Ratio Uncertainty | Weighted standard deviation (weights from uncertainties below) | 9% |
| | NLLS fit parameters for 9432 Å integration | 3 – 7%[a] |
| | NLLS fit parameters for 15874 Å integration | 6 – 15%[a] |
| | Blackbody calibration integration | 2% |
| | Blackbody Calibrator drift and temperature uncertainty | 3% |
| 9432 Å Transition Probability Uncertainty | Wood et al. 2013 | 17% |
| Final 15874 Å Transition Probability Uncertainty | Line strength ratio and 9432 Å transition probability uncertainties combined in quadrature | 19% |

[a]Varies due to the S/N of the individual spectra.

References

Bizzarri, A., Huber, M. C. E., Noels, A., et al. 1993, A&A, 273, 707

Cirasuolo, M., Afonso, J., Bender, R., Bonifacio, P., Evans, C., et al. 2011, ESO Messenger 145, 11

Eisenstein, D. J., Weinberg, D. H., Agol, E., et al. 2011, AJ, 142, 72

Hinkle, K., Wallace, L., & Livingston, W. 1995, "Infrared Atlas of the Arcturus Spectrum 0.9 – 5.3 µm" (Astronomical Society of the Pacific: San Francisco CA)

Hodapp, K. W., Hora, J. L., Hall, D. N. B., et al. 1996, New Astronomy, 1, 177

Ives, D., & Bezawada, N. 2007, Nuclear Instruments and Methods in Physics Research A, 573, 107

Kurucz, R. L., & Bell, B. 1995, Atomic Line Data CD-ROM No. 23 (Smithsonian Astrophysical Observatory: Cambridge, MA)

Peck, E. R., & Reeder, K. 1972, JOSA, 62, 958

Saloman, E. B. 2012, JPCRD, 41, 013101

Thorne, A., Litzén, U., & Johansson, S. 1999, in Spectrophysics: Principles and Applications (Berlin: Springer-Verlag)

Wilson, C. J., Hearty, F., Skrutskie, M. F., et al. 2010, Proceedings of the SPIE, 7735, 77351C

Wilson, C. J., Hearty, F., Skrutskie, M. F., et al. 2012, Proceedings of the SPIE, 8446, 84460H

Wood, M. P., Lawler, J. E., Sneden, C., & Cowan, J. J. 2013, ApJS, 208, 27

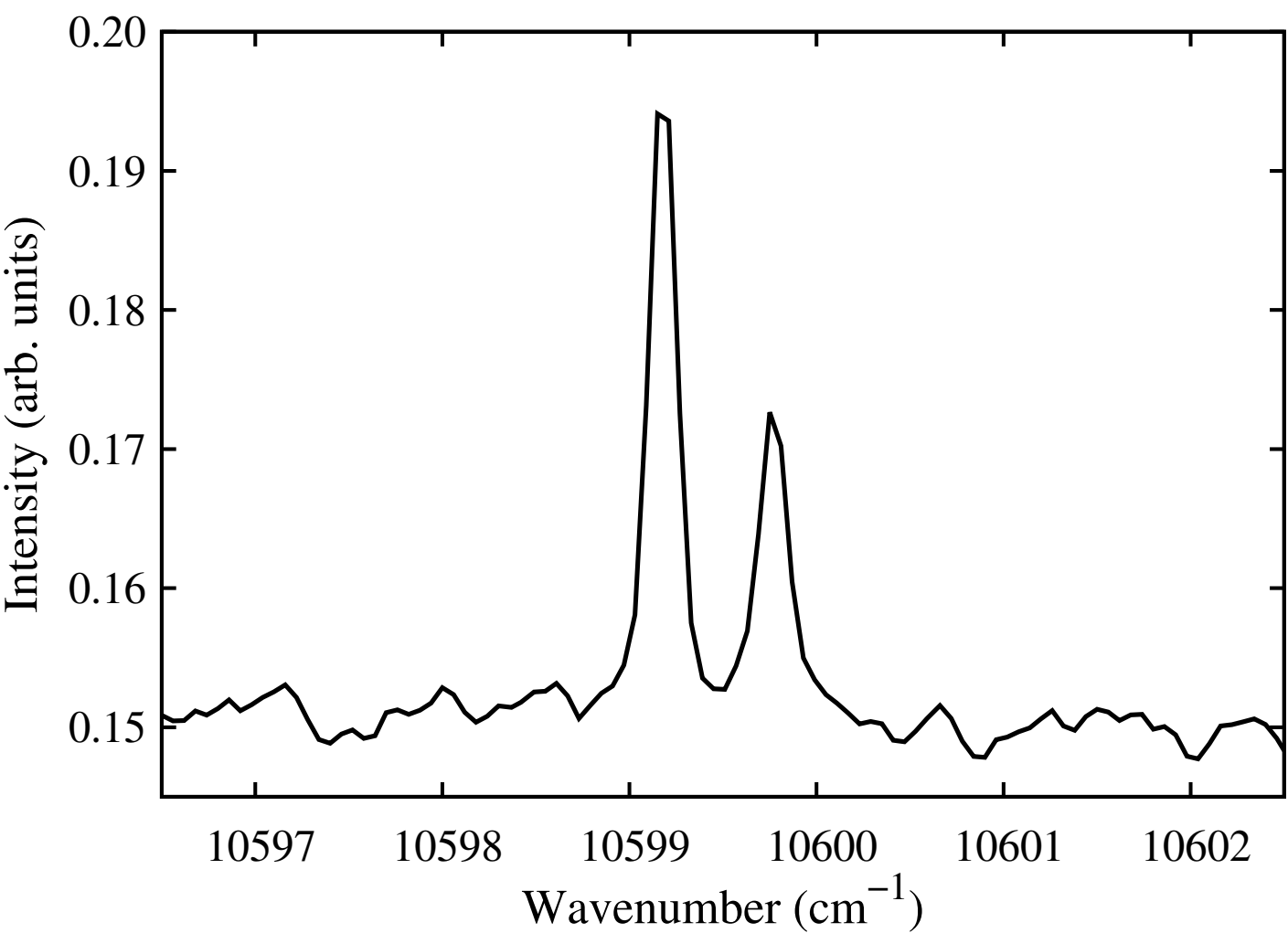

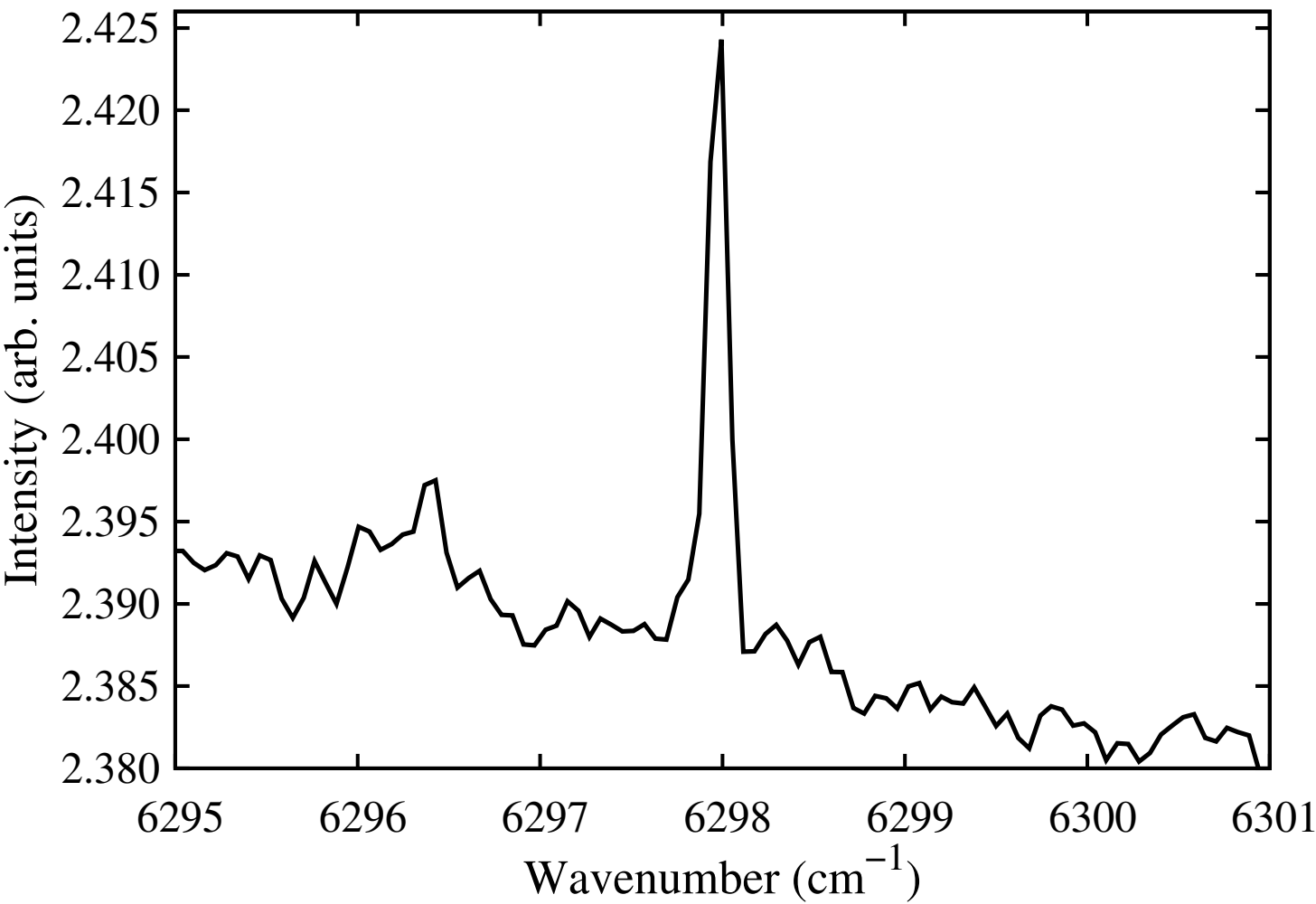